\documentclass[12pt]{article}
\usepackage{epsfig}

\def\be{\begin{equation}}
\def\ee{\end{equation}}
\def\ba{\begin{array}{c}}
\def\ea{\end{array}}

\def\ben{$$}
\def\een{$$}
\begin{document}

\titlepage

 \begin{center}{\Large \bf

%${\cal PT}-$symmetric localization of free waves

${\cal PT}-$symmetric knotting of coordinates: a new, topological
mechanism of quantum confinement.

 }\end{center}

\vspace{5mm}

 \begin{center}
Miloslav Znojil

\vspace{3mm}

\'{U}stav jadern\'e fyziky\footnote{e-mail: znojil@ujf.cas.cz }
 AV \v{C}R, 250 68 \v{R}e\v{z}, Czech
Republic

\end{center}

\vspace{5mm}

\section*{Abstract}

%
%Quantum knots:
%
%${\cal PT}-$symmetric potentialless bound states

In Quantum Mechanics, the so called ${\cal PT}-$symmetric bound
states $\psi(r)$ can live on certain nontrivial contours ${\cal
C}$ of complex coordinates $r$. We construct an exactly solvable
example of this type where Sturmian bound states exist in the
absence of any confining potential. Their origin is purely
topological: our ``quantum knot" model uses certain specific,
spiral-shaped contours ${\cal C}^{(N)}$ which $N-$times encircle a
logarithmic branch point of $\psi(r)$ at $r=0$.

 \vspace{9mm}

\noindent
 PACS 03.65.Ge, 11.10.Kk, 11.30.Na, 12.90.+b

%\vspace{9mm}

 \begin{center}
%{\small \today, hanketobo.tex file}
%=\psi^{(phys)}(x) \in
%L_2(-\infty,\infty)=E^{(phys)}
\end{center}

\newpage

\section{Introduction \label{s1} }

For any spherically symmetric single-particle Hamiltonian
$H=-\triangle +V(|\vec{x}|^2)$ acting in ${\it I\!\!\!L}^2({\it
I\!\!\!R}^D)$ let us recollect the related radial Schr\"{o}dinger
equation
 \be
  -\frac{d^2}{dr^2}\,\psi (r)+ \frac{\ell(\ell+1)}{r^2}
  \,\psi (r)+ V(r^2)
  \,\psi (r)= E \,\psi (r)\,,
  \ \ \ \ \ \
  r \in {I\!\!\!R}^+\,
   \label{SEnotfree}
 \ee
with $\ell=(D-3)/2+ m$ in the $m-$th partial wave. At $D=1$ one
only needs $m=0$ (for the even parity states) and $m=1$ (for the
odd parity states) while at all the higher dimensions $D \geq 2$,
the sequence of the angular-momentum indices $m$ in
eq.~(\ref{SEnotfree}) is infinite, $m=0, 1, \ldots$.

At any dimension $D \geq 1$ one can shift the coordinate $r  \
\longrightarrow \ r_\epsilon = s - {\rm i}\,\epsilon$ using a
constant $\epsilon > 0$ and a new real variable $s\in {I\!\!\!R}$
in eq.~(\ref{SEnotfree}). As far as we know, the resulting change
of the Hamiltonian $H^{(radial)} \ \longrightarrow \
H^{(shifted)}$ has been first proposed by Buslaev and Grecchi
\cite{BG} who choose their parity-symmetric real potential
$V(r^2)$ in a very specific anharmonic-oscillator form. They
proved that their manifestly non-Hermitian new Hamiltonian
$H^{(shifted)}\neq \left (H^{(shifted)}\right )^\dagger$ can be
perceived as physical since it is {\em strictly isospectral} to a
one-dimensional double-well oscillator model with manifestly
Hermitian and safely confining Hamiltonian $Q=Q^\dagger$.
Marginally, they also noticed (cf. their Remark~4, {\it loc.
cit.}) that $H^{(shifted)}$ is ${\cal PT}-$symmetric, with ${\cal
T}$ denoting the complex conjugation operator and with ${\cal P}$
representing the standard operator of parity. Unfortunately, the
latter Remark passed virtually unnoticed \cite{VGpriv}. Only five
years later, Bender with coauthors \cite{BM,BB,BBjmp} proved much
more successful in re-attracting attention to the ${\cal
PT}-$symmetric family of the bound-state models where, as they
conjectured, the spectrum of the energies has a very good chance
of remaining real and observable.

In the latter context let us recollect our recent studies
\cite{tob,tobscatt,tob2} and, in  their spirit, replace the
Buslaev's and Grecchi's complex integration contour
 \ben
  {\cal C}={\cal C}^{(BG)} = \{r =
  s-{\rm i} \epsilon\,|\, \epsilon> 0\,,
  \,s \in I\!\!R\}\,
 \een
by an element ${\cal C}^{(N)}$ of a family of its generalizations
which will be specified below (cf. section \ref{druzka}). In order
to simplify the construction (cf. section \ref{soudruzka}) and the
discussion (cf. sections \ref{drzka}), we shall
only consider the simplified, dynamically trivial model with
vanishing $V(r^2)$,
 \be
  -\frac{d^2}{dr^2}\,\psi (r)+ \frac{\ell(\ell+1)}{r^2}
  \,\psi (r)= E \,\psi (r)\,,
  \ \ \ \ \ \
  r \in {\cal C}^{(N)}\,.
   \label{SEfree}
    \label{SEto}
    \label{SEr}
 \ee
In order to extend the scope of such a model slightly beyond its
purely kinematical version (cf. also the summary in section
\ref{ruch}), it is easy to add an interaction term $V(r) =
\gamma/r^2$. Then, the implicit definition of the effective $\ell$
in (\ref{SEr}),
 \be
 \ell(\ell+1) = \gamma + \left (m+\frac{D-3}{2}
 \right )\,\left (m+\frac{D-1}{2}
 \right )\,,\ \ \ \ \ \ \
 m=0, 1, \ldots
 \ee
enables us to treat our $\gamma$ or $\ell=\ell(\gamma)$ as a
continuous real parameter. Under this assumption we intend to show
that in spite of absence of any confining force, our
Schr\"{o}dinger eq.~(\ref{SEfree}) defined along topologically
nontrivial integration paths generates Sturmian bound states $\psi
(r)\in {\it I\!\!\!L}^2\left ({\cal C}^{(N)} \right ) $ at certain
discrete sets of $\gamma$ and/or $\ell=\ell(\gamma)$.

%\newpage
%
%\newpage
%********** Figure 1 zde
\begin{figure}[h]                     %instead of \begin{figure}[t]
\begin{center}                         %instead of \begin{center}
\epsfig{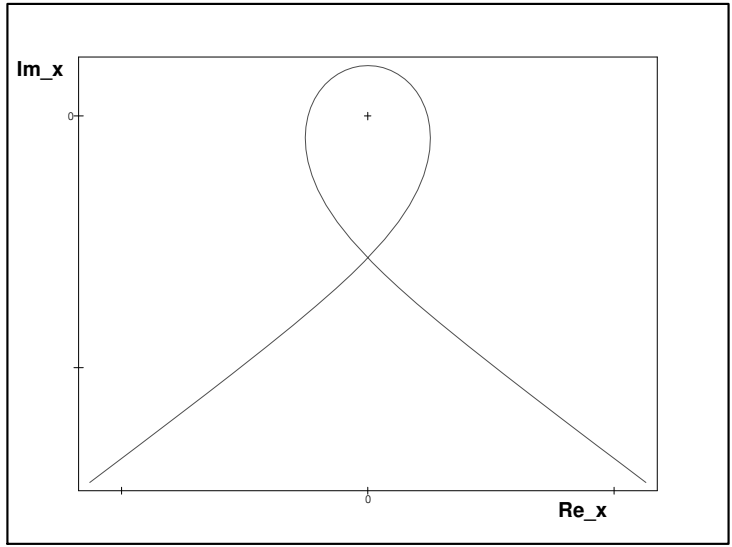}
\end{center}                         %instead of \end{center}
\vspace{-2mm} \caption{Sample of the curve ${\cal C}^{(N)}$ with
$N=1$.
 \label{fione}}
\end{figure}
%\newpage

%
%\newpage
%
%\newpage
%********** Figure 2 zde
\begin{figure}[h]                     %instead of \begin{figure}[t]
\begin{center}                         %instead of \begin{center}
\epsfig{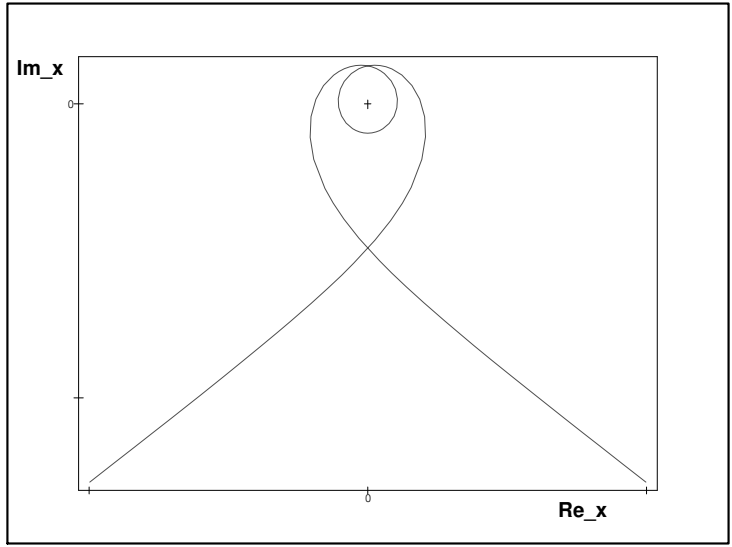}
\end{center}                         %instead of \end{center}
\vspace{-2mm} \caption{Sample of the curve ${\cal C}^{(N)}$ with
$N=2$.
 \label{fitwo}}
\end{figure}
%\newpage

\section{Integration paths
${\cal C}^{(N)}$ \label{druzka} }

In certain complex domains of $r$, the general solution of any
ordinary linear differential equation of the second order can be
expressed as a superposition of some of its two linearly
independent solutions,
 \be
 \psi(r) = c_1\,\psi^{(1)}(r) +c_2\,\psi^{(2)}(r)\,.
 \label{ansa}
 \ee
In our present class of models (\ref{SEfree}) let us distinguish
between the subdomains of the very small, intermediate and very
large $|r|$. In the former case we can choose
 \be
 \psi^{(1)}(r) = r^{\ell+1} + corrections\,,\ \ \ \
 \psi^{(2)}(r) = r^{-\ell} + corrections\,,\ \ \ \
 \ \ \
 \ \ \ |r| \ll 1\,,
 \label{mala}
 \ee
while we would prefer another option in the latter extreme, with
$\kappa=\sqrt{E}$ in
 \be
 \psi^{(1,2)}(r) =
 \exp \left ({\pm {\rm i}\,\kappa\,r} \right )
 + corrections\,,\ \ \ \
 \ \ \ |r| \gg 1\,.
 \label{asymptotics}
 \ee
In between these two regimes, our differential eq.~(\ref{SEfree})
is smooth and analytic so that we may expect that all its
solutions (\ref{ansa}) are also {\em locally} analytic.

In the vicinity of the origin $r=0$ our differential equation has
a pole. For the time being let us assume that our real parameter
$\ell$ in eq.~(\ref{SEfree}) is irrational. In such a case, both
the components of our wave functions (\ref{ansa}) (as well as
their arbitrary superpositions) would behave, {\em globally}, as
multivalued analytic functions defined on a certain multisheeted
Riemann surface ${\cal R}$. In the other words, our wave functions
possess a logarithmic branch point in the origin, i.e. a branch
point with an infinite number of Riemann sheets connected at this
point \cite{BpT}.

Separately, one might also study the simplified models with the
rational $\ell$s which correspond to the presence of an algebraic
branch point at $r=0$ which would connect a finite number of
sheets \cite{BpT}. Let us briefly mention here just the simplest
possible scenario of such a type where $\ell(\ell+1)=0$ to that
either $\ell=-1$ or $\ell=0$. In such a setting, eq.~(\ref{ansa})
just separates $\psi(r)$ into its even and odd parts so that the
Riemann surface itself remains trivial, ${\cal R}\, \equiv
\,l\!\!\!C$.

In the generic case of a multisheeted  ${\cal R}$ we intend to
show that the asymptotically free form of our differential
eq.~(\ref{SEfree}) with the independent solutions
(\ref{asymptotics}) {\em can} generate bound states. One must
exclude, of course, the contours running, asymptotically, along
the real line of $r$ since, in such a case, {\em both} our
independent solutions $\psi^{(1,2)}(r)$ remain oscillatory and
non-localizable. The same exclusion applies to the parallel,
horizontal lines ${\cal C}^{(BG)}$ in the complex plane of $r$. In
the search for bound states, both the ``initial" and ``final"
asymptotic branches of our integration paths ${\cal C}^{(N)}$ must
have the specific straight-line form $r=\pm |s|\,e^{{\rm
i}\varphi}$ with a non-integer ratio $\varphi/\pi$. Thus, we may
divide the asymptotic part of the complex Riemann surface of $r\in
{\cal R}$ into the sequence of asymptotic sectors
 \begin{equation}
 %\bullet
%  \ \ \ \ \ \ \ \ \ \ \ \ \ \
 {\cal S}_0 = \{r=-{\rm
i}\,\varrho\,e^{{\rm i}\,\varphi}\,|\,\varrho \gg 1\,,\ \varphi
\in (-\pi/2,\pi/2)\} \label{hola}\,.
 \ee
\be
 %\bullet
%  \ \ \ \ \ \ \ \ \ \ \ \ \ \
   {\cal S}_{\pm k} = \{r=-{\rm i}\,e^{\pm {\rm
i}\,k\,\pi}\,\varrho\,e^{{\rm i}\,\varphi}\,|\,\varrho \gg 1\,,\
\varphi \in (-\pi/2,\pi/2)\},\ \ \ \ k = 1, 2, \ldots\,.
\label{holas}
 \ee
We are now prepared to define the integration contours ${\cal
C}^{(N)}$. For the sake of convenience, we shall set all their
``left" asymptotic branches ${\cal C}^{(left)}$ in the same sector
${\cal S}_{0}$ and specify $r=\left (s+s_0\right )\,(1+{\rm
i}\varepsilon)$ where $s \in (-\infty, -s_0)$, $\varepsilon>0$ and
$s_0>0$. The subsequent middle part of ${\cal C}^{(N)}$ must make
$N$ counterclockwise rotations around the origin inside ${\cal R}$
while $s \in (-s_0,s_0)$. Finally, the ``outcoming" or ``right"
asymptotic branch of our integration contour ${\cal C}^{(N)}$ with
$s \in (s_0, \infty)$ must lie in another sector ${\cal S}_{2N}$
of ${\cal R}$, i.e., in the Riemann sheet where the requirement of
${\cal PT}-$symmetry \cite{tobscatt} forces us to set $r=\left (
s-s_0\right )\,(1-{\rm i}\varepsilon)$. In this spirit, our
illustrative Figures \ref{fione} -- \ref{fithree} sample the
choice of $N=1$, $N=2$ and $N=3$, respectively.

%********** Figure 3 zde
\begin{figure}[h]                     %instead of \begin{figure}[t]
\begin{center}                         %instead of \begin{center}
\epsfig{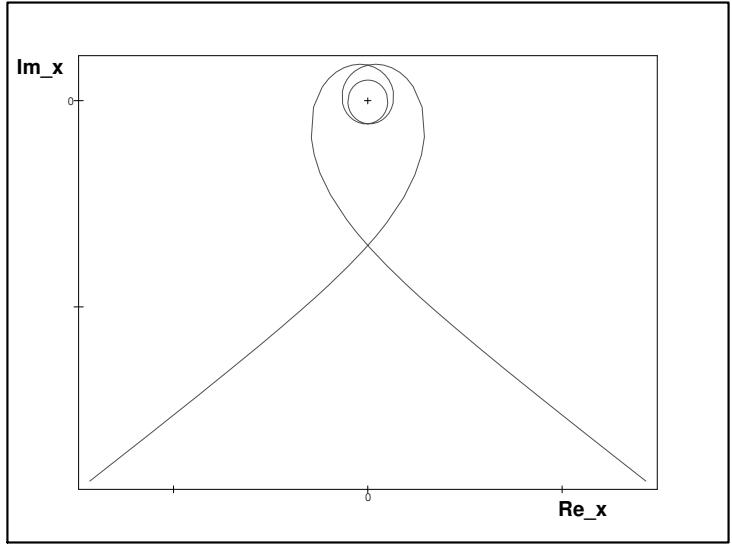}
\end{center}                         %instead of \end{center}
\vspace{-2mm} \caption{Sample of the curve ${\cal C}^{(N)}$ with
$N=3$.
 \label{fithree}}
\end{figure}
%\newpage

\section{Bound states along nontrivial paths \label{soudruzka}}

It remains for us to impose the asymptotic boundary conditions
requiring that our wave functions vanish at $s \to \pm \infty$. As
long as our integration path ${\cal C}^{(N)}$ performs $N$
counterclockwise rotations around the origin, this form of the
asymptotic boundary conditions will already guarantee the
normalizability of our bound-state wave functions $\psi (r)\in
{\it I\!\!\!L}^2\left ({\cal C}^{(N)} \right ) $ (cf. the similar
situation encountered in the models with confining
potentials~\cite{BG,tob,Alvarez}).

In the subsequent step  we may set $z=\kappa r$ and
$\psi(r)=\sqrt{z}\,\varphi(z)$ in our bound-state problem
(\ref{SEto}) with the (by assumption, real) $E=\kappa^2$. This
reduces eq.~(\ref{SEto}) to the Bessel differential equation with
the pair of the two well known independent special-function (say,
Hankel-function~\cite{Ryzhik}) solutions which may be inserted in
our ansatz (\ref{ansat}),
 \be
 \psi(r) = c_1\,\sqrt{r}\,H^{(1)}_\nu(\kappa\,r)
 +c_2\,\sqrt{r}\,H^{(2)}_\nu(\kappa\,r)\,,\ \ \ \  \nu =
 \ell+1/2\,.
 \label{ansat}
 \ee
At $|{\rm arg}\,z| < \pi$ and ${\rm Re}\,\nu > -1/2$, the
asymptotics of its components are given by the respective formulae
8.451.3 and 8.451.4 of ref.~\cite{Ryzhik},
 \ben
 \sqrt{\frac{\pi z}{2}}\,H^{(1)}_\nu(z)=
 \exp\left [{\rm i}\left (z-\frac{\pi(2\nu+1)}{4}
 \right )\right ]\,\left (1-\frac{\nu^2-1/4}{2{\rm i} z} + \ldots
 \right )
 \,,
 \een
 \ben
 \sqrt{\frac{\pi z}{2}}\,H^{(2)}_\nu(z)=
 \exp\left [-{\rm i}\left (z-\frac{\pi(2\nu+1)}{4}
 \right )\right ]\,\left (1+\frac{\nu^2-1/4}{2{\rm i} z} + \ldots
 \right )
 \,.
 \een
This implies that inside the even-subscripted sectors ${\cal
S}_{2k}$ our ansatz (\ref{ansat}) combines the asymptotically
growing (and, hence, unphysical) component $H^{(1)}_\nu(z)$ with
the asymptotically vanishing and normalizable, physical component
$H^{(2)}_\nu(z)$. {\it Vice versa}, in all the odd-subscripted
sectors ${\cal S}_{2k+1}$ we would have to eliminate, in
principle, the asymptotically growing $H^{(2)}_\nu(z)$ and to keep
the asymptotically vanishing $H^{(1)}_\nu(z)$.

We may start our discussion of the existence of the localized
bound states from the straight-line contour ${\cal C}={\cal
C}^{(BG)} = {\cal C}^{(0)}$ which is all contained in the zeroth
sector ${\cal S}_0$. This immediately implies that with ${\rm
Im}\,r \ll -1$, the asymptotically vanishing solution
 \ben
 \psi^{(1)}(r) =
 \sqrt{ r}\,H^{(2)}_\nu(\kappa\,r)
 =const\,\cdot\,\exp (-{\rm i}\,\kappa\,r)+ corrections\,
 \een
remains unconstrained at all the real $\kappa$. Obviously, the
spectrum remains non-empty and bounded from below. This means that
the low-lying states remain stable with respect to a random
perturbation. A less usual feature of such a model is that its
energies densely cover all the real half-line $I\!\!R^+$. This
feature is fairly interesting {\em per se}, although a more
detailed analysis of its possible {\em physical} consequences lies
already beyond the scope of our present brief note.

Our eigenvalue problem becomes not too much more complicated when
we turn attention to the spiral- or knot-shaped integration
contours ${\cal C}^{(N)}$ with~$N> 0$. In such a case,
fortunately, the exact solvability of our differential equation
enables us to re-write ansatz (\ref{ansa}) in its fully explicit
form which remains analytic on all our Riemann surface ${\cal R}$.
Once we choose our ``left" asymptotic sector as ${\cal S}_{0}$,
the ``left" physical boundary condition fixes and determines the
acceptable solution on the initial sheet,
 \be
 \psi(r) = c\, \sqrt{r}\,H^{(2)}_\nu(\kappa r)\,,\ \ \ \ \
 r \in {\cal S}_0\,.
 \label{BS}
 \ee
After the $N$ counterclockwise turns of our integration path
${\cal C}^{(N)}$  around the origin this solution gets transformed
in accordance with formula 8.476.7 of ref.~\cite{Ryzhik} which
plays a key role also in some other solvable models \cite{CJT},
 \be
 H^{(2)}_\nu\left (ze^{{\rm i}m\pi}\right )=
 \frac{\sin (1+m)\pi\nu}{\sin \pi \nu}\,
 H^{(2)}_\nu(z)+
 e^{{\rm i}\pi\nu}\,
 \frac{\sin m\pi\nu}{\sin \pi \nu}\,
 H^{(1)}_\nu(z)\,.
 \label{nuit}
 \ee
Here we have to set $m=2N$. This means that the existence of a
bound state will be guaranteed whenever we satisfy the ``right"
physical boundary condition, i.e., whenever we satisfy the
elementary requirement of the absence of the unphysical component
$H^{(1)}_\nu(z)$ in the right-hand side of eq.~(\ref{nuit}).

The latter requirement is equivalent to the doublet of conditions
 \be
  2N\nu = integer\,,\ \ \ \ \ \nu \neq integer\,.
 \ee
This means that at any fixed and positive value of the energy
$E=\kappa^2$ and at any fixed winding number $N=1,2,\ldots$, our
present quantum-knot model generates the series of the bound
states at certain irregular sequence of angular momenta avoiding
some ``forbidden" values,
 \be
 \ell =\frac{M-N}{2N}\,,\ \ \ \ \
 M =  1, 2, 3, \ldots\,,\ \ \ \ \
 M \neq 2N, 4N, 6N, \ldots\,.
 \label{formu}
 \ee
These bound states exist and have the analytically continued
Hankel-function form (\ref{BS}) {\em if and only if} the {\em
kinematical input} represented by the angular momenta $\ell$ is
{\em restricted to the subset} represented by formula
(\ref{formu}).

Our construction is completed. Once we restrict our attention to
the purely kinematic model with $\gamma=0$, we can summarize that
at the odd dimensions $D=2p+1$ giving $\ell=n+p-3/2$ we may choose
any index $n$  and verify that formula (\ref{formu}) can be read
as a definition of the integer quantity $M=(2n+2p-1)\,N$ which is
not forbidden. At the even dimensions $D=2p$ we equally easily
verify that the resulting $M$ is always forbidden so that our
quantum-knot bound states do not exist at $V(r)=0$ at all.

The latter dichotomy appears reminiscent of its well-known
non-quantum real-space analogue, but the parallel is misleading
because in quantum case the freedom of employing an additional
coupling constant $\gamma$ enables us to circumvent the
restrictions. Indeed, once we select {\em any} dimension $D$,
angular-momentum index $m$, winding number $N$ and any ``allowed"
integer $M$, our spectral recipe (\ref{formu}) may simply be
re-read as an explicit definition of the knot-supporting value of
the coupling constant
 \ben
 \gamma=\left (\frac{M}{2N}
 \right )^2-
 \left (m+\frac{D-2}{2}
 \right )^2\,.
 \een
This implies that at non-vanishing $\gamma$s, the quantum knots
can and do exist at all dimensions.

\section{Discussion \label{drzka}}

\subsection{Complexified coordinates}

In the language of physics, our present construction and solution
of a new and fairly unusual exactly solvable quantum model of
bound states is based on the freedom of choosing the knot-shaped,
{\em complex} contours of integration ${\cal C}$. This trick is
not new \cite{BG,BB} and may be perceived as just a consequence of
the admitted loss of the observability of the coordinates in
PT-symmetric Quantum Mechanics \cite{Carl}.

From an experimentalist's point of view, the omission of the
standard assumption that the coordinate ``should be" an observable
quantity, $x \in I\!\!R$, is not entirely unacceptable since the
current use of the concept of quasi-particles paved the way for
similar constructions. Related Hamiltonians could be called, in
certain sense, manifestly non-Hermitian. Still, they are currently
finding applications in nuclear physics (where they are called
quasi-Hermitian \cite{Geyer}). The loss of the reality of the
coordinates is also quite common in field theory where the similar
unusual Hamiltonians are being rather called CPT-symmetric
\cite{BBJ} or crypto-Hermitian \cite{Smilga}.

In a pragmatic phenomenological setting, the fairly unusual nature
of the new structures of spectra seems promising. At the same
time, the formalism itself is now considered fully consistent with
the standard postulates of quantum theory. In the language of
mathematics, the emergence of its innovative features may be
understood as related to the non-locality in $x$, i.e., to the
replacement of the standard scalar product
 \ben
 \langle \psi\,|\,\phi\rangle = \int \psi^*(x) \phi(x) dx
 \een
by its generalized, nonlocal modifications \cite{Carl,workshops}
 \ben
 \langle \psi\,|\,\phi\rangle = \int \psi^*(x)\,
 \Theta(x,y)\,\phi(y) dx \, dy\,.
 \een
Although this leaves an overall mathematical consistency and
physical theoretical framework of Quantum Theory virtually
unchanged \cite{Carl}, a new space is being open, {\it inter
alii}, to the topology-based innovations. In principle, they might
inspire new developments of some of the older successful
applications of the formalism ranging from innovative
supersymmetric constructions \cite{susy} to cosmology
\cite{Alicos}, occasionally even leaving the domain of quantum
physics \cite{my}.

\subsection{Towards a simplification of mathematics}

The very existence of the present knot-type model is certainly
based on nontrivial, non-Dirac scalar-product Hilbert-space
kernels $\Theta \neq I$ (which is our present symbol for the
quantity denoted as $T$ in ref.~\cite{Geyer} or as $\eta_+$ in
refs.~\cite{Ali} or as a factorized symbol ${\cal CP}$ in
ref.~\cite{BBJ} or ${\cal PQ}$ in \cite{sigma} etc). This being
said, a part of the price for an enhancement of the scope of the
theory lies in the necessity of the explicit study of the
equivalence transformations between the {\em simple} quantum model
(which is non-Hermitian and ${\cal PT}-$symmetric) and its
``physical", Hermitian alternative representation(s)
$H^{(Hermitian)}$. This transformation is not only non-local and
complicated but also strongly Hamiltonian-dependent.

One could try to re-establish some parallels with the {\em
exceptional} Buslaev's and Grecchi's oscillators \cite{BG} where
$H^{(Hermitian)}$ happened to be, incidentally, {\em still} of a
local, differential-equation nature. In the similar fortunate
cases (cf. also ref.~\cite{Cali} for another illustration), the
construction of $H^{(Hermitian)}$ remained reducible to the
construction of a certain effective interaction $V^{(eff)}(r) \in
I\!\!R$.

Partially, a similar technique of simplification can be also
applied to our present model with $V=0$. Indeed, although it seems
that all the variations of the spectrum must be attributed to the
mere kinematical aspects of the model in question, one could
employ the philosophy of ref. \cite{tob} to imagine that the
nonequivalent effective interactions $V^{(eff)}(r)$ could be also
attributed to the nonequivalent knot-shaped integration contours
${\cal C}^{(N)}$. An implementation of such a perspective would
rely upon the conformal mapping
 \be
 r = -{\rm i}\,e^{{\rm i}\,x/2}\,.
 \label{mappi}
 \ee
It simply maps each Riemann sheet in $r$ on a stripe in complex
$x-$plane where $\Re e\ x$ plays the role of the angular
coordinate while $\Im m\ x$ simulates the radial coordinate. In
the other words, the polar-coordinates-resembling
mapping~(\ref{mappi}) can be used to compress several sheets of a
multisheeted Riemann surface ${\cal R}$ onto the single reference
complex plane of the auxiliary complex variable~$x$. In this
manner an effective potential would emerge quite naturally (cf.
also ref.~\cite{CJT} in this respect).

%
%The radial straight lines in the complex $r-$plane will correspond
%to a constant $\Re e\ x$ [specified in eqs. (\ref{hola}) and
%(\ref{holas})]. In the other words, the asymptotics of the
%contours ${\cal C}$ in $r-$plane can be visualized as maps of
%certain vertical lines in the auxiliary $x-$plane. In contrast, a
%concentric circle in $r$ emerges, at a fixed value of $\Im m\ x$,
%as a map of a horizontal line in the complex $x-$plane. Thus, any
%circular part of our tobogganic path ${\cal C}^{(N)}$ which
%encircles the origin in the plane of $r$ becomes tractable as a
%map of a sufficiently long horizontal line in the complex plane of
%$x$. This assures us that the rightward motion in the $x-$plane
%can be understood as a rotation around the origin in the
%$r-$plane. In this process,

\section{Outlook \label{ruch} }

%{Towards a re-introduction of potentials $V \neq 0$}

We can summarize that  our present example belongs to such a type
of quantum theory where

\begin{itemize}

\item  ``coordinates" $r$ may  complexify and lose their immediate
observability,

\item the topological structure of paths ${\cal C}$ can often be
simplified via suitable conformal changes of variables
\cite{tob,tobscatt,tob2},

\item  a ``suitable" \cite{Geyer,BBJ,Ali,Quesne} redefinition of
the inner product in the Hilbert space is needed,

\item  complex potentials  $V(r) \in l\!\!\!C$ are allowed.

\end{itemize}

 \noindent
Let us add, finally, a few remarks on the latter, nontrivial and
appealing point. Firstly, one must imagine that our present,
exactly solvable  $N>1$ quantum-knot bound-state problem would
become purely numerical after its immersion in virtually any
external confining potential. Secondly, the survival of the
reality of the new bound-state spectra (i.e., of the measurability
of the energies) would have to be {\em individually} secured or
proved as the necessary condition of the physical acceptability of
the models with $V(r) \neq 0$.

The latter point has already been properly emphasized by Bender
and Boettcher \cite{BB} who choose the one-parametric family of
the one-dimensional power-law forces $
 V^{(BB)}(x) = x^2({\rm i}\,x)^\delta\,
$ and supported the conjecture $E \in I\!\!R$ by the persuasive
numerical and semiclassical arguments. The more rigorous (and
unexpectedly difficult)  proof only followed several years later
\cite{DDT}. It is worth noticing that the latter proof found {\em
also} its most natural formulation in the general $D-$dimensional
setting with $D>1$ and/or with the variable real $\ell$.

One of the merits of working with the power-law-dominated
confining potentials was that the asymptotic behavior of the
related wave functions can be analyzed easily (cf. also refs.
\cite{BG,BB} and \cite{Alvarez} where such an analysis has been
performed and discussed in detail). One of the less expected
results of these studies (revealed, tested and verified also by
several other groups of authors \cite{BBjmp,CJT,mytri} during 1998
- 1999) has been the observation of the explicit
contour-dependence of the spectra. In fact, this discovery, so
important in our present context, is even older since its first
indications can even be traced back to the pioneering 1993 letter
by Bender and Turbiner~\cite{BT}. In our papers
\cite{tob,tobscatt} we emphasized and extended this observation by
contemplating the ``usual" potentials in combination with the
``unusual", topologically nontrivial, spiral-shaped contours
${\cal C}={\cal C}^{(N)}$ of complex coordinates.

On this background, the exact solvability of the present,
force-free example came as a surprise. It could be perceived as an
important new illustration of the changes of the energy spectra
mediated by the use of the non-equivalent integration contours.
Still, in the nearest future, a more numerically-oriented return
to the more realistic problem of the contour-shape-dependence of
the spectra generated by some nontrivial  ${\cal PT}-$symmetric
potentials in eq.~(\ref{SEr}) seems unavoidable.

\section*{Acknowledgements}

Supported by the GA\v{C}R grant Nr. 202/07/1307, by the M\v{S}MT
``Doppler Institute" project Nr. LC06002 and by the NPI
Institutional Research Plan AV0Z10480505.

%\newpage

\section*{Figure captions}

\subsection*{Figure 1. Sample of the curve ${\cal C}^{(N)}$ with
$N=1$.}

\subsection*{Figure 2. Sample of the curve ${\cal C}^{(N)}$ with
$N=2$.}

\subsection*{Figure 3. Sample of the curve ${\cal C}^{(N)}$ with
$N=3$.}

\end{document}